# DC measurements of macroscopic quantum levels in a superconducting qubit structure with a time-ordered meter

D.S. Crankshaw, K. Segall, D. Nakada, T.P. Orlando[a], L.S. Levitov[b], S. Lloyd[c]; S.O. Valenzuela, N. Markovic, M. Tinkham[d]; K.K. Berggren[e]

[a]*Department of Electrical Engineering and Computer Science, MIT, Cambridge, MA 02139*

[b]*Department of Physics, MIT, Cambridge, MAs 02139*

[c]*Department of Mechanical Engineering, MIT, Cambridge, MA 02139*

[d]*Department of Physics, Harvard University, Cambridge, MA 02138*

[e]*MIT Lincoln Laboratory, Lexington, MA 02420*

Abstract: DC measurements are made in a superconducting, persistent current qubit structure with a time-ordered meter. The persistent-current qubit has a double-well potential, with the two minima corresponding to magnetization states of opposite sign. Macroscopic resonant tunneling between the two wells is observed at values of energy bias that correspond to the positions of the calculated quantum levels. The magnetometer, a Superconducting Quantum Interference Device (SQUID), detects the state of the qubit in a time-ordered fashion, measuring one state before the other. This results in a different meter output depending on the initial state, providing different signatures of the energy levels for each tunneling direction. From these measurements, the intrawell relaxation time is found to be about 50 μs.

## I. Introduction

The study of macroscopic quantum effects in superconductors is motivated both by interest in the extension of quantum mechanics to the classical world [1] and by the possibility of constructing a quantum information processor [2]. Macroscopic quantum effects, such as resonant tunneling [3], quantum superposition states [5,6], and time-dependent coherent oscillations [4-8] have recently been observed. In these experiments, measurements were made on charge [4], flux [5,6], and current [7,8].

One particular superconducting system that has been under study is the persistent-current qubit (PC qubit), a superconducting ring interrupted by three Josephson junctions [9]. When an external magnetic flux bias near one-half of a flux quantum ($\Phi_0 = h/2e$) is applied, it has two stable classical states of electrical current circulating in one direction or the other, resulting in measurable opposing magnetizations. It can be modeled as a double-well potential in a three-dimensional potential landscape (one dimension for each of the junction's phase variables, or three other variables which span the space), where the minimum of each well corresponds to one these two magnetization states. Depending on the parameters, the system may have multiple quantum energy levels in one of the two wells, where each level has approximately the same magnetization. Energy levels in a similar system, the radio-frequency superconducting quantum interference device (rf SQUID), have been measured by studying resonant tunneling between the two wells [3]. Experiments on an rf SQUID have used a separate, damped SQUID magnetometer as the meter. This approach gives a continuous readout of the magnetization, but also couples unwanted dissipation into the system.

In a recent paper, we showed how coupling an underdamped dc SQUID magnetometer to a PC qubit resulted in time-ordered measurements of the two states, where one state is observed before the other [10]. In those experiments we studied the classical, thermally driven regime of operation. In the present paper we detail the effects of a time-ordered meter on the dc measurements of the PC qubit in the quantum regime. The quantum levels are detected by observing resonant tunneling between the two wells. The positions agree well with calculations of the qubit energy band structure, and the energy bias of level repulsions indicate where tunneling occurs between the two wells. While the PC qubit has inherent symmetry between the two states, the time ordering of the measurements causes an asymmetry in the meter output, which depends on the initial state of the qubit. We demonstrate this asymmetry, and also show how the meter shifts the positions of the energy levels as a function of the external flux bias and the SQUID current bias. Finally, by measuring the width and height of the tunneling peaks as a function of the SQUID ramp rate, we find a fitted value of the intra-well relaxation of order microseconds.

## II. Qubit Parameters and Measurement Process

The qubit and dc SQUID are both fabricated at Lincoln Laboratory in a Niobium trilayer process [13]. The circuit diagram is shown in Figure 1. The qubit consists of a superconducting ring interrupted by three Josephson junctions, two of which are designed to have the same critical current, $I_c$, and the third of which has a critical current of $\alpha I_c$, where $\alpha$ is less than 1. The measurement dc SQUID, which surrounds the qubit, has two equal Josephson junctions with critical currents of $I_{c0}$, where $I_{c0}>I_c$. The PC qubit loop is 16x16 $\mu m^2$ in area, and the dc SQUID is 20x20 $\mu m^2$ in area, with self-inductances of about $L_q = 30\,\text{pH}$ and $L_S = 60\,\text{pH}$ respectively. They have a mutual inductance of approximately $M = 25\,\text{pH}$. These inductances are calculated using FastHenry [14], then refined through experimental measurements of the SQUID's response to magnetic field, as explained in Appendix A. The critical current density is 370 A/cm$^2$, and the critical current of the SQUID junctions are measured to be $I_{c0} = 5.3\,\mu\text{A}$, consistent with an area of 1.4 $\mu m^2$. $I_c$ and $\alpha$ can be determined experimentally from our previous thermal activation studies, which give $\alpha = 0.63$ and $I_c = 1.2\,\mu\text{A}$ [10]. These values are within the range of estimated values from the process parameters.

By changing the magnetic flux through the superconducting loop, the depth of each well of the double-well potential changes, with one becoming deeper as the other becomes shallower. The energy bias ($\varepsilon$) is the energy difference between minima of the two wells. (We will also use it to indicate the difference between energy levels in opposite wells, using a subscript to indicate which energy levels we are measuring the difference between.) It is periodic with frustration, $f_q$, which is the magnetic flux bias of the qubit in units of flux quanta. At $f_q = 0.5$, the depth of the two wells are equal, and the energy bias varies nearly linearly with frustration, such that $\varepsilon$ is approximately $4\pi\alpha E_J\left(f_q - 0.5\right)$, where $E_J = I_c\Phi_0/2\pi$ is the Josephson energy of each of the two larger junctions of the qubit.

At low temperatures, thermal activation is insufficient to overcome the barrier between the two wells when $f_q = 0.5$. In this case, hysteresis is observed, where the PC qubit remains in the state in which it is prepared until it is measured, even though this state is no longer the minimum energy state. The qubit is prepared in a state by changing its magnetic flux bias to where the system has a single well and allowing the qubit to relax to its ground state. Then it is brought to the magnetic flux bias where it is measured. The qubit will remain in the state where it was prepared, either the left well (the **0** state) or the right well (the **1** state) until it has the opportunity to escape to the opposite well.

## III. Energy Level Structure Calculations

The energy level diagram in Figure 2 includes in its calculations the phases of all three junctions. If the inductance of the qubit is small enough, the phase of the three junctions is confined by flux quantization, and only two independent variables are necessary in the Hamiltonian. The requirement for this approximation is that $\beta_L = L_q / L_J < 0.01$ [16], where $L_q$ is the inductance of the qubit loop and $L_J$ is the Josephson inductance of each of the larger junctions. This is not the case in our sample, where $\beta_L = 0.01$. In order to correctly solve the Hamiltonian of our device, we need to include the inductance and solve for the three-dimensional Hamiltonian. We start by making a change of variables from the phases of the three junctions to $\Theta_1$ and $\Theta_2$, which are node phases, and $I_m$, which is the current around the PC qubit loop (we also use the variable $I_p$ to denote the persistent current in the qubit, but $I_p$ is technically the expectation variable in each state, while $I_m$ is the quantum variable, thus $I_p = \langle I_m \rangle$). This gives us the equalities in Equation (1) for converting the phase variables of the junctions into $\Theta_1$, $\Theta_2$, and $I_m$:

$$\begin{aligned}
\varphi_1 &= \Theta_1 - \left(\frac{1-b}{2}\right) 2\pi \left(\frac{L_q I_m}{\Phi_0} + f_q\right) \\
\varphi_2 &= \Theta_2 + \left(\frac{1-b}{2}\right) 2\pi \left(\frac{L_q I_m}{\Phi_0} + f_q\right) \\
\varphi_3 &= \Theta_2 - \Theta_1 - (b) 2\pi \left(\frac{L_q I_m}{\Phi_0} + f_q\right)
\end{aligned} \qquad (1)$$

The variable $b$, which describes how the self-inductance of the qubit is divided among its branches, is arbitrary so long as it is less than one. We can define its value as $1/(1+2\alpha)$ so that it eliminates any product terms of the time derivative of $I_m$ and the time derivative of either $\Theta_1$ or $\Theta_2$ in the Hamiltonian. By changing variables again, this time to $\Theta_+ = (\Theta_1 + \Theta_2)/2$ and $\Theta_- = (\Theta_1 - \Theta_2)/2$, while defining the effective masses associated with these two variables as $M_+ = 2(\Phi_0 / 2\pi) 2C_j$ and $M_- = (2 + 4\alpha)(\Phi_0 / 2\pi) 2C_j$, where $\underline{C_j}$ is the junction capacitance, we get the Hamiltonian in Equation (2).

$$H = \frac{1}{2} M_{+} \dot{\Theta}_{+}^{2} + \frac{1}{2} M_{-} \dot{\Theta}_{-}^{2} + \frac{\alpha}{2+4\alpha} C_{J} L_{q}^{2} \dot{I}_{m}^{2} +$$
$$E_J \left[ 2 + \alpha - 2\cos\Theta_{+} \cos\left( \Theta_{-} - \left( \frac{1-b}{2} \right) 2\pi \left( \frac{L_q I_m}{\Phi_0} + f_q \right) \right) - \right. \quad (2)$$
$$\left. \alpha \cos\left( -2\Theta_{-} - (b) 2\pi \left( \frac{L_q I_m}{\Phi_0} + f_q \right) \right) \right]$$

While complex, this is numerically solvable by discretizing the variables $\Theta_+$, $\Theta_-$, and $I_m$ into $\Theta_{+_i}, \Theta_{-_j}$, and $I_{m_k}$ respectively, and creating a Hamiltonian matrix whose elements are $H_{pq} = \left\langle \Theta_{+_i} \Theta_{-_j} I_{m_k} \,|\, \hat{H} \,|\, \Theta_{+_r} \Theta_{-_s} I_{m_t} \right\rangle$, where ***p*** and ***q*** are indices that map onto all the permutations of ***i,j,k*** and ***r,s,t*** respectively. ***H*** is a square matrix where each side has a length equal to the product of the number of discretized elements of $\Theta_p$, $\Theta_m$, and $I_m$. The matrix must be kept sparse in order to solve on a computer due to memory limitations, and the band structure in Figure 2 shows the eigenvalues of this Hamiltonian matrix as the external magnetic flux bias is changed. The inclusion of self-inductance changes the energy band diagram, most significantly by reducing the level repulsion, since the barrier between the two wells is greater due to the need to overcome the qubit's self inductance. Each avoided level crossing with a measured signature in the band diagram is labeled, using **a**, **b**, **c**, **c***′*, **d**, and **e** for the level crossings when $f_q$ is less than 0.5, and using **A**, **B**, **C**, **C***′*, **D**, and **E** when $f_q$ is greater than 0.5. Although the energy scales are such that some of the avoided level crossings appear to actually cross, there is a small amount of energy level repulsion even at $f_q = 0.5$. There are multiple energy levels in each well, and each level crossing corresponds to the alignment of the lowest level in one well and one of the energy levels in the other well, as is shown in the double well potentials in Figure 2. This results in two eigenstates, a symmetric state and an antisymmetric state spanning both wells, with an energy difference equal to the level splitting shown in the energy band diagram.

IV. Results and Discussion

To determine the state of the PC qubit, we ramp the electrical current in the dc SQUID until it switches to the voltage state. The measuring dc SQUID remains in the zero-voltage state as long as the current through it is below the switching current; when it passes this current it develops a finite voltage. The switching current depends on the magnetic flux through the SQUID, and has a maximum value of $I_{sw0} = 2I_{c0}\left|\cos(\pi f_S')\right|$, where $I_{c0}$ is the critical current of each of the two SQUID junctions and $f_S'$ is the total magnetic flux through the SQUID in units of flux quanta. Since the qubit's two states have different magnetizations, the two states have different switching currents, $\mathbf{I_0}$ for state **0** and $\mathbf{I_1}$ for state **1**. The switching of the SQUID to the voltage state is a stochastic process with a variance that is measurable but significantly smaller than the signal we are measuring. The ramp rate is typically 4 μA/ms, which means that it takes 125 μs to ramp from $\mathbf{I_0}$ to $\mathbf{I_1}$. If the qubit is in the **0** state, the SQUID switches as soon as it arrives at $\mathbf{I_0}$. If it is in the **1** state and remains there, the SQUID does not switch until it reaches $\mathbf{I_1}$.

Figure 3 illustrates the observed hysteresis of the qubit. Part (a) of this figure shows the probability of measuring the qubit in each state by plotting $P_\mathbf{1}$-$P_\mathbf{0}$, where $P_\mathbf{1}$ is the probability of finding it in state **1** and $P_\mathbf{0}$ is the probability of finding it in state **0**. The solid curve is the probability measurement when the system is prepared in the **1** state, while the dashed curve shows the measurement when it is prepared in the **0** state. This measurement occurs at ~20 mK bath temperature, where the thermal energy is $k_BT$~1.7 μeV. Part (b) of this figure shows how the difference between the 50% point of the two curves shown in (a) varies with temperature. We call this difference the "hysteresis loop width." The width is constant up to $T = 200\,\text{mK}$, then decreases linearly as temperature increases. Thermal activation causes the qubit to change state once the barrier is on the order of $k_BT$, and the barrier height changes linearly with magnetic flux bias. It intercepts zero at 550 mK, or 47 μeV.

The hysteresis loop width closes as the thermal activation increases since the magnetic flux bias which gives a significant escape rate for thermal activation is closer to $f_q = 0.5$ as the temperature increases. Thermal activation has the rate of $\Gamma_{th} = (7.2\Delta U\omega_0/2\pi Q k_B T)\exp(-\Delta U/k_B T)$. Calculations of the potential energy, confirmed by previous experiments, show how the barrier between the two states, $\Delta U$, varies with the magnetic flux bias of the qubit [10]. Near $f_q = 0.5$, the barrier for the qubit to transition from the **1** to the **0** state can be written as $\Delta U_{\mathbf{10}}(f_q) \approx \Delta U(0.5) + 2\pi\alpha E_J(f_q - 0.5)$, where $2\pi\alpha E_J = 9500\mu\,\text{eV}$. $\Delta U(f_q = 0.5)$ is $2\pi\alpha E_J\left[2\cos^{-1}\left(2\sqrt{(1-\alpha)^2/3}\right) - \cos^{-1}\left(\sqrt{(1-\alpha)^2/3\alpha^2}\right)\right]$, or about 210 μeV [10], or 2.4 K. $\Delta U_{\mathbf{01}}(f_q) \approx \Delta U(0.5) - 2\pi\alpha E_J(f_q - 0.5)$ is the barrier for the transition from **0** to **1** at the same flux bias. The thermal activation rate which results in the qubit falling to the lowest energy point rather than remaining in the local minimum should be approximately a constant, and since the exponential is the greatest influence on $\Gamma_{th}$, $\Delta U(f_q)/k_BT$ should be a constant which we designate ξ. To calculate the slope, we take $210\mu\,\text{eV} + 9500\mu\,\text{eV}(f_q - 0.5) = \xi(86\mu\,\text{eV/K})T$, which can be differentiated by $dT$ to give that $df_q/dT = 0.009\xi\,\text{K}^{-1}$. This is the location of one side of the hysteresis loop, while the total width is twice this, such that $dw/dT = 2\,df_q/dT = 0.018\xi\,\text{K}^{-1}$. The measured slope is 0.131 $\text{K}^{-1}$ in Figure 3(b), which corresponds to a ξ of 7.3.

Figure 4 shows the number of switching events at various values of current and magnetic flux bias. The horizontal axis represents the externally applied magnetic flux to the SQUID in terms of frustration, while the vertical axis corresponds to the current bias of the SQUID. The shading indicates the number of switching events that occur at each point in the external flux bias and current bias coordinates. Experimentally, about $10^3$ measurements are taken at each value of external flux bias, so each vertical slice represents a histogram of these measurements. Over most of the parameter space this figure shows that two preferred states exist, corresponding to the **1** and **0** states of the qubit. These states create two "lines" across the figure, reminiscent of the results in [12].

However, the detailed signatures of the switching events when the qubit is prepared in the **1** state (Fig 4(a)) differ from when it is prepared in the **0** state (Fig 4(b)), even though the energy biases are mirror images of each other around $f_q = 0.5$ as shown by the double-well potentials drawn above Figure 3(a). Figure 4(a) shows stripes in the region in-between the two lines of switching currents whereas Figure 4(b) has no switching events in this in-between region. The two lines of switching currents in Figure 4(b) show island-like regions whereas Figure 4(a) does not. The plots in Figure 3, which are derived from the same data, also reflect this asymmetry. Both Figures 4(a) and 4(b) show a range of flux bias where both states can be measured. One important reason for the difference between the two plots is the path, in bias current and external magnetic field, followed by the SQUID, which is illustrated by the dashed line in the two figures. Rather than being completely vertical, indicating a ramp in SQUID current while the external magnetic field is held constant, the external magnetic field also changes during the current ramp due to the preparation of the qubit state. This, and the influence of the time-ordered measurements, results in differences in the data due to macroscopic quantum tunneling.

We first consider the data in Figure 4(a), where the system is initially prepared in the **1** state. When the qubit is prepared in the **1** state, it will remain there as long as the bias is such that the local minimum exists, unless it has some mechanism to escape this local minimum, such as thermal activation or macroscopic quantum tunneling. At 15 mK bath temperature, thermal activation is effectively frozen out, and macroscopic quantum tunneling is the dominant process; therefore, one expects to see tunneling at the locations of the energy level crossings on the left side of the band diagram in Figure 2. The probability that the state will transition from **1** to **0** depends on the tunneling rate and the time that the system remains in the level-crossing region. Note that if the qubit transitions to the **0** state before the SQUID bias current reaches $I_1$ but after the current is past $I_0$, the SQUID switches immediately, and we record a switching event in the region between $I_0$ and $I_1$. One might expect vertical stripes to appear between $I_0$ and $I_1$ at the external flux biases corresponding to the level crossings. The observed stripes are instead curved because the total flux bias of the qubit, $f_q$, depends on flux coupled from the readout SQUID as well as the externally applied flux, $f_q^{ext}$. The total flux biasing the qubit is $f_q = f_q^{ext} + MI_{cir}/\Phi_0$, where $f_q^{ext}$ is the externally applied flux bias, $M$ is the mutual inductance between the dc SQUID and the qubit, and $I_{cir}$ is the circulating current in the SQUID. The circulating current in the dc SQUID decreases as the bias current increases. The circulating current is calculated from the Josephson equations of a SQUID with a finite self-inductance to be $I_{cir} = I_{c0}\sin(\pi f_S')\sqrt{1 - I_{bias}^2/(2I_{c0}\cos(\pi f_S'))^2}$, where $f_S'$ is the effective flux bias of the SQUID, which has a form similar to $f_q$, specifically $f_S' = f_S + MI_p/\Phi_0 + L_S I_{cir}/\Phi_0$. Here $f_S$ is the externally applied flux bias to the dc SQUID and $L_S$ is the self-inductance of the SQUID. $I_p$ is the persistent current in the qubit, which is nearly constant at $\alpha I_c = 760\,\text{nA}$, and whose sign depends on the state in which the qubit is prepared. The appearance of $I_{cir}$ in $f_S'$ requires that the circulating current be solved self-consistently. This calculation shows that the qubit's effective flux bias approaches the externally applied flux as it moves closer to the switching current. The lines of constant effective flux bias are drawn on Figure 4(a), using the same values of the mutual inductance that we derive in Appendix A, and the stripes in the switching events match up with these lines of constant effective flux bias ($f_q$). Furthermore, the lines of $f_q$ that match up to the stripes indicate the effective flux biases where the level crossings occur in the qubit, and these stripes compare well to the calculated level crossings of the qubit (shown in Figure 2) based on the parameters we obtained from thermal activation experiments [11]. By ramping the bias current more slowly, the system spends more time near the level crossings that have a smaller tunneling rate and they show up more clearly; in this way, all the level crossings have been mapped out and show all the expected energy levels [15]. The effect of the ramp rate is discussed further below.

We now consider the data in Figure 4(b), where the qubit is initially prepared in the **0** state. If the qubit remains in the **0** state, then the SQUID switches to the voltage state at $I_0$. The qubit cannot change from the **0** state to the **1** state after $I_0$, since by that point the SQUID will already have switched. If it transitions from **0** to **1**, this must occur before the qubit reaches $I_0$. There are no observed switching events when the current bias is between $I_0$ and $I_1$, indicating that when it makes the transition from **0** to **1** it does not transition back. This is expected if after tunneling into one of the higher energy levels of the right well, it relaxes to a lower energy state where it is no longer in alignment with the energy level of the shallow well. Slowing down the measurement has a noticeable effect when the qubit is prepared in the **0** state, as shown in Figure 6, which plots the relative probability of finding the qubit in each state as $P_1$-$P_0$. The peaks grow larger while maintaining their position, indicating that the probability of transition grows due to the slowing of the SQUID ramp rate. This suggests that the transition rate from the **0** to **1** state is comparable to the time over which the levels are near alignment in the SQUID ramp rate. If the rate were much faster, then all the population would tunnel to the **1** state. If it were much slower, then none of the population would tunnel.

Recall that the flux bias of the qubit is $f_q = f_q^{ext} + M\, I_{cir}/\Phi_0$. Since $I_{cir}$ changes as the SQUID is ramped, and $f_q^{ext}$ is pulsed when the qubit is prepared, $f_q$ is a function of time. The direction of the magnetic field pulse depends on what state the qubit is prepared in, but the change in $I_{cir}$ is in the same direction as long as the hysteresis loop width is small compared to the SQUID periodicity. Thus, when prepared in the **1** state, the two factors sum, while preparation in the **0** state causes the two factors to oppose one another. Since both the state preparation and the coupled dc SQUID field are nonlinear and they only happen simultaneously for a short length of time, they do not completely cancel out. They do, however, balance for roughly 50 μs, where the flux bias of the qubit is nearly constant. This is seen in Figure 5, which maps out the value of $f_q$ as it changes over time. $f_q$ plateaus briefly during the current ramp when the qubit is prepared in the **0** state. If this plateau corresponds to an energy bias where there is a high rate of quantum tunneling between the wells, this results in a strong probability of tunneling which gives a sharp peak in the probability data ($P_1$-$P_0$).

## V. Simulation of transitions

To simulate the effect of the SQUID current bias ramp on the state of the qubit, an equation for the tunneling rate is needed. These measurements resemble those taken by Lukens [3], and described theoretically by Averin [17] to give an equation for the rate of transition from the lowest energy level in one well to a high energy level in the other:

$$\tau_{\mathrm{i}}^{-1} = \frac{\Delta_i^{\,2}\Gamma_i}{2\Delta_i^{\,2} + \Gamma_i^2 + 4\varepsilon_{\mathrm{i}}^{\,2}} \tag{3}$$

where $\tau_i^{-1}$ is the transition rate for level crossing $i$, $\Delta_i$ and $\varepsilon_i$ are the tunnel splitting and energy bias of the specific level crossing respectively, and $\Gamma_i$ is the rate at which the qubit relaxes from the high energy level in the deeper well to one of the lower energy levels. The transition is not considered complete without this decay, which prevents the phase from returning to the original state. This state is at a lower energy at both the beginning and end of our measurement (although not during the entire course of it), and thus is energetically more favorable than the highest energy state in the other well. $\Delta_i$ is a function of the quantum model of the qubit, and can be calculated from the parameters that we already know. $\varepsilon_i$, the energy bias, is the energy difference between the energy levels in each well. This is equal to $4\pi\alpha E_J(f_q - f_i)$, where $f_i$ is the position, in magnetic flux bias, of the individual level crossing we are considering. We will consider the $\Gamma_i$ to be a constant for each level crossing. This relaxation to the lower energy levels is the fitting parameter, with the guideline that the higher the energy level, the more quickly it should relax. Running a simulation of the transition probability as $f_q$ changes during the SQUID current bias ramp gives Figure 6(a) and (b), showing a match between the theoretical model and the experiment at two different ramp rates. Using this model with theoretically calculated numbers for $\Delta_i$, the values of $\Gamma_i$ range from $(1\ \mu\mathrm{s})^{-1}$ to $(100\ \mu\mathrm{s})^{-1}$ for all the levels, which is a long decay time for intrawell relaxation. It should be noted that the theory allows some trade-off between $\Delta_i$ and $\Gamma_i$, so that a smaller $\Delta$ would correspond to a faster relaxation time. Environmental fluctuations may effectively decrease $\Delta_i$ while increasing $\Gamma_i$, implementing this trade-off. While we modeled this reduction in $\Delta_i$ using Wilhelm's formulation [18], the exact amount is necessarily uncertain since it depends on the total environment of the qubit, which we cannot directly observe.

There are several strong peaks in this data, including two that are right next to each other. In the quantum model of the qubit, the only point that would give two level-crossings so close together would be due to a transverse mode of the three-dimensional well, which produces an energy level of the first excited state in the $\Theta_+$ direction near the energy level as the second excited state in the $\Theta_-$ direction. We presume that we are able to observe this mode only because of an asymmetry in the two larger junctions, due to fabrication variances, which produce a coupling between the transverse mode energy level in the deep well and the lowest state in the shallow well.

Using both preparation states, we have observed energy levels within each well, which are separated from the ground state by frequencies of 28 GHz, 53 GHz, 60 GHz, and 72 GHz. This is measured from the location of the stripes when prepared in the **1** state and the location of the peaks when prepared in the **0** state, using the estimation that $\varepsilon \approx 4\pi\alpha E_J\left(f_q - 0.5\right)$. The locations of these level crossings agree well with the energy band diagram in Figure 2, which is calculated by numerically finding the eigenstates of the Hamiltonian.

VI. Summary

In summary, we have observed macroscopic quantum tunneling in persistent current qubits. The observed stripes when the qubit is prepared in the **1** state, even more than the distinct variations in the state populations when it is prepared in the **0** state, indicate quantum level crossings of states in the qubit's wells. We can use these observed stripes and variations to refine our parameters in the quantum simulation so that it gives the same crossings. The measurements have mapped out energy levels in the double-well potential that agree well with the energy level calculation. Experiments are underway to observe coherent oscillations between the states.

Acknowledgments

The authors would like to thank B. Singh, J. Lee, J. Sage, E. Macedo, and T. Weir for experimental help, and L. Tian and W. D. Oliver for useful discussions. The authors would also like to thank the staff of the MIT Lincoln Laboratory Microelectronics Laboratory for their assistance in sample fabrication. This work is supported in part by the AFOSR grant F49620-01-1-0457 under the DoD University Research Initiative on Nanotechnology (DURINT) and by ARDA. The work at Lincoln Laboratory was sponsored by the Department of Defense under the Department of the Air Force contract number F19628-00-C-0002. Opinions, interpretations, conclusions and recommendations are those of the authors and not necessarily endorsed by the Department of Defense.

[1] A.J. Leggett and A. Garg, *Phys. Rev. Lett.* **54**, p. 857, 1985.
[2] S. Lloyd, *Science* **261**, p. 1569, 1193.
[3] R. Rouse, S. Han, and J.E. Lukens, *Phys. Rev. Lett.* **75**, p. 1614, 1995.
[4] Y. Nakamura, Y.A. Pushkin, J.S. Tsai, *Nature* **398**, p. 786, 1999.
[5] J.R. Friedman, V. Patel, W. Chen, S.K. Tolpygo, and J.E. Lukens, *Nature* **406**, p. 43, 2000.
[6] C.H. van der Wal, A.C.J. ter Haar, F.K. Wilhelm, R.N. Schouten, C.J.P.M. Harmans, T.P. Orlando, S. Lloyd, and J.E. Mooij, *Science* **290**, p. 773, 2000.
[7] J.M. Martinis, S. Nam, J. Aumentado, and C. Urbina, *Phys. Rev. Lett.* **89**, p. 117901, 2002.
[8] Y. Yu, S. Han, X. Chu, S. Chu, and Z. Wang, *Science* **296**, p. 889, 2002.
[9] J. E. Mooij, T. P. Orlando, L. Levitov, L. Tian, C. H. van der Wal, and S. Lloyd, *Science* **285**, p. 1036, 1999.
[10] K. Segall, D. Crankshaw, D. Nakada, T.P. Orlando, L.S. Levitov, S. Lloyd, N. Markovic, S.O. Valenzuela, M. Tinkham, K.K. Berggren, *Phys. Rev. B* **67**, 220506, 2003.
[11] While [10] accurately gave us the values of $E_J$, $\alpha$, and $Q$, the value of $E_C$ was not as well determined. These measurements show us the location of the level crossings in Figure 4(a). We can determine the value of $E_C$ which gives these biases for the level crossings, which we find to be 3 μeV.
[12] H. Takayanagi, H. Tanaka, S. Saito, and H. Nakano, *Physica Scripta* **T102**, p. 95,

2002.
[13] K.K. Berggren, E.M. Macedo, E.M. Feld, J.P. Sage, *IEEE Trans. Appl. Supercond.* **9**, p. 3271, 1999.
[14] M. Kamon, M.J. Tsuk, J.K. White, *IEEE Trans. on Microwave Theory and Techniques* **42**, p. 1750, 1994.
[15] D.S. Crankshaw, *Measurement and On-chip Control of a Niobium Persistent Current Qubit*, Ph.D. Thesis, MIT, 2003.
[16] D.S. Crankshaw and T.P. Orlando, *IEEE Trans. on Appl. Supercond.* **11**, p. 1006, 2000.
[17] D.V. Averin, J. R. Friedman, and J.E. Lukens, *Phys. Rev. B***,** vol. 62, p. 11802, 2000.
[18] F.K. Wilhelm, *Phys. Rev. B* **67,** 060503, 2003.

Appendix A

Calculating the mutual inductance between the dc SQUID and the qubit is straight forward. The self-inductance of the SQUID can be determined from the transfer function of magnetic flux to switching current. From these values, we can calculate the circulating current in the SQUID as it varies with frustration. The shape of this curve, especially its minimum and any bimodal features due to multiple wells in its potential, tells us the value of $\beta_{L,S} = L_S / L_{J,S}$ . Figure 7(a) shows the periodicity with which the qubit step appears in the SQUID transfer function. Since the SQUID is 1.53 times the size of the qubit, the step should appear at every 1.53 periods in the SQUID curve. This periodicity arises because, while both the SQUID and the qubit have a periodicity of $\Phi_0$, the SQUID receives more flux due to its larger size. However, the qubit is not perfectly periodic, as is shown in Figure 7(b), where the points mark the difference between the qubit step's position and where it would appear if it occurred with perfect periodicity, $\Delta f_q^{ext}$. This deviation indicates that there are sources of magnetic field other than that applied by the external magnet, the strongest of which is the field coupled to the qubit by the circulating current in the dc SQUID. (The SQUID is also influenced by the circulating current in the qubit, but since this is only one-seventh of the value of the circulating current in the SQUID, it can be safely neglected.) The total field seen by the qubit is $f_q = f_S / 1.53 + M I_{cir} / \Phi_0$ , where $f_S$ is the frustration of the SQUID from the externally applied field and $I_{cir}$ is the circulating current in the SQUID. Thus $\Delta f_q^{ext} = f_q - f_S / 1.53 = M\, I_{cir} / \Phi_0$ . If we use a least squares fit to find a value of $M$ that causes $MI_{cir}/\Phi_0$ to intersect the $\Delta f_q$ data points, we can solve for $M$, which we find to be about 25 pH. This produces the curve in Figure 7(b) that intersects the data points.

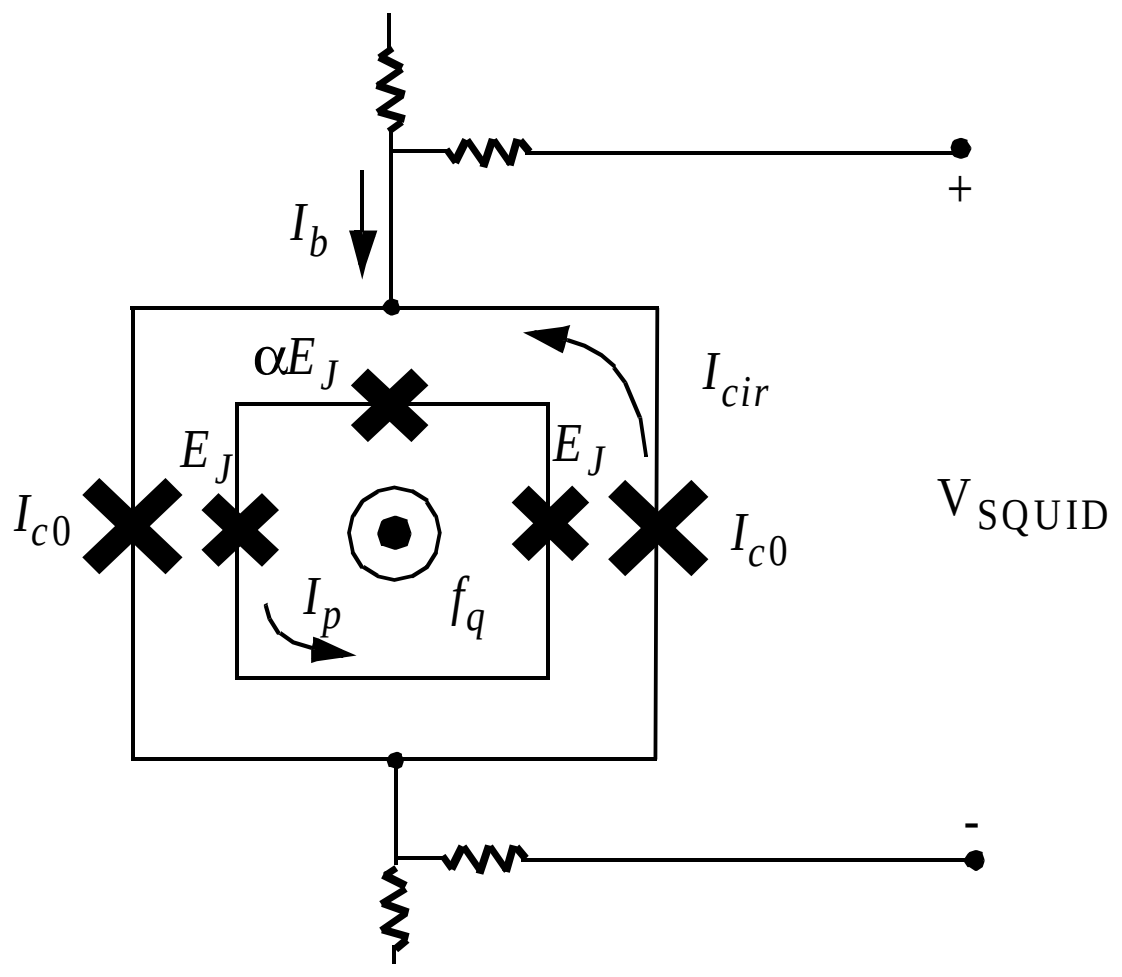

(a)

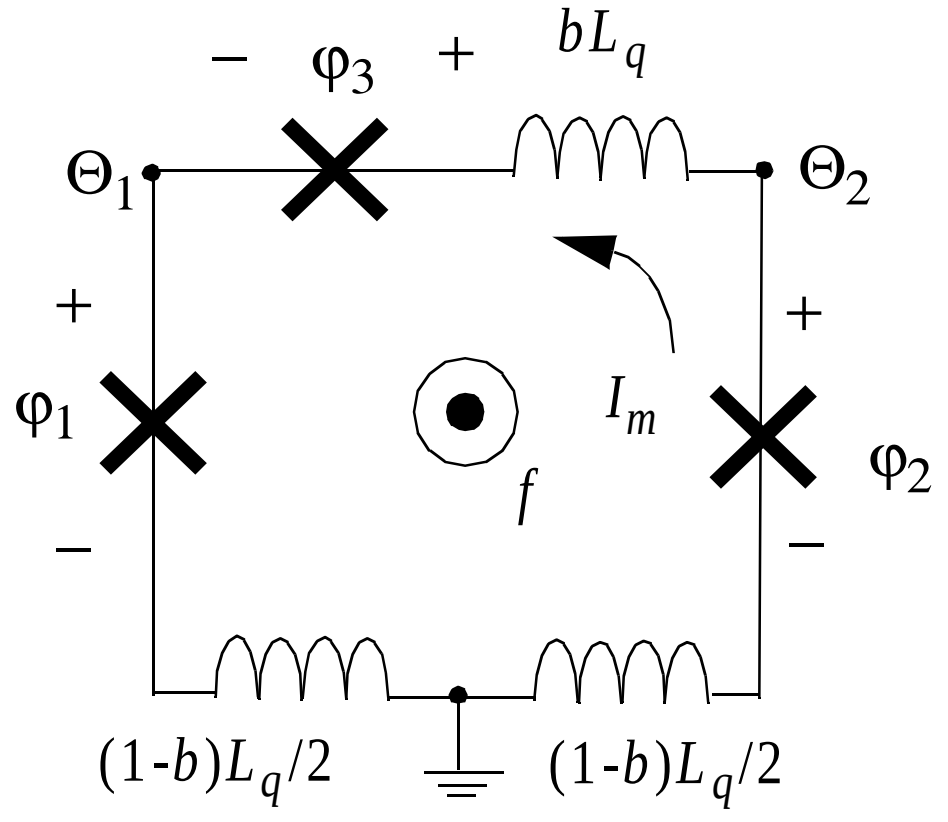

(b)

FIGURE 1. (a) A circuit diagram of the qubit structure, a three junction loop, and the two junction SQUID. (b) The circuit diagram used to derive the quantum mechanical model of the qubit. The inductance is distributed among the branches, the inductance on the branch of the smallest junction having a value of $bL_q$, while the inductances on the other two branches share a value of $(1\text{-}b)L_q$. The node phases, $\Theta_1$ and $\Theta_2$, are shown in the figure.

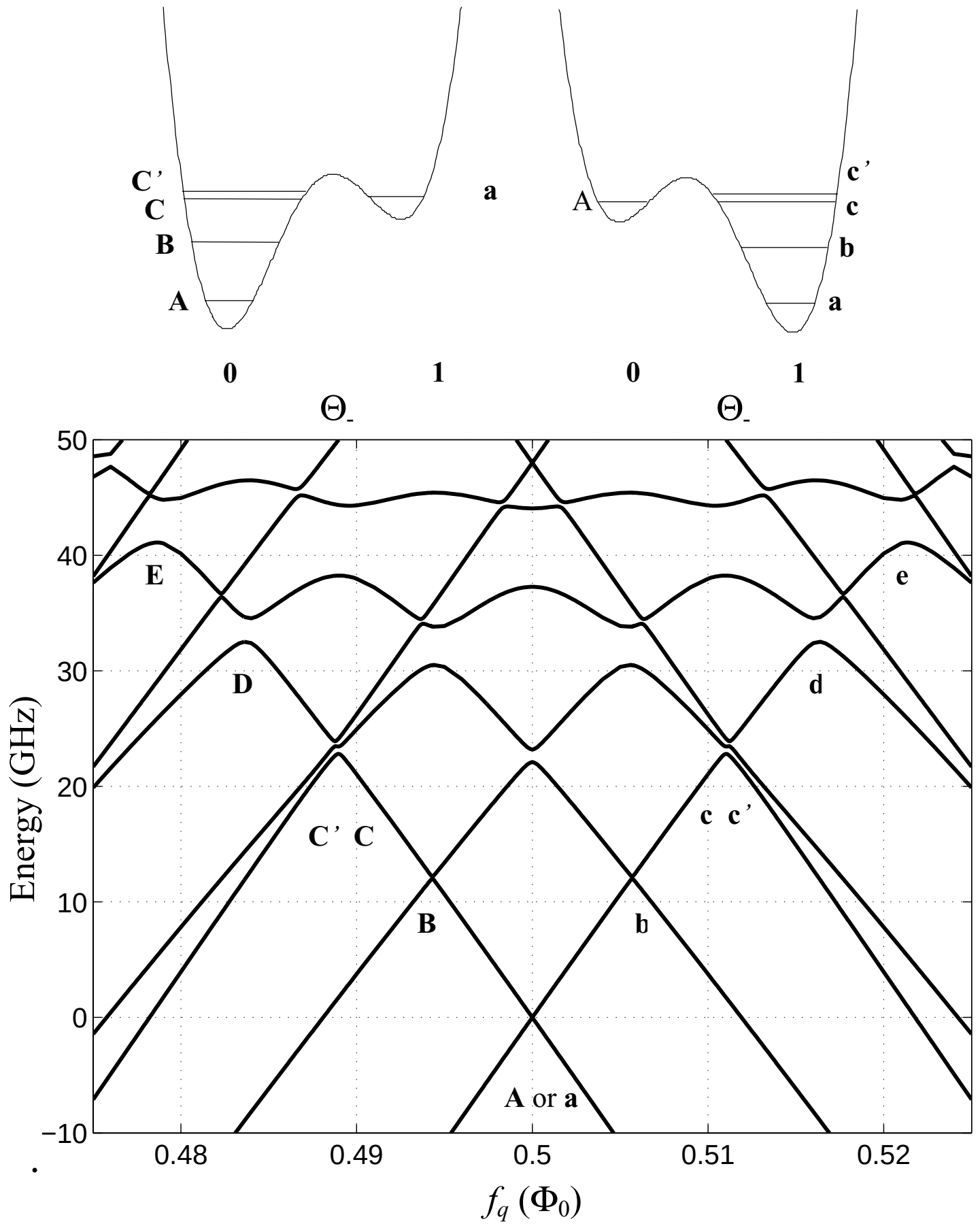

FIGURE 2. An energy band diagram of the PC qubit with the parameters described in the text, with the magnetic flux bias of the qubit in units of flux quanta ($f_q$) as the horizontal axis. The transitions between the **0** and **1** states occur at the avoided level crossings. These are at $f_q$=0.478, 0.483, 0.487, 0.488, 0.494, and 0.500 on the left side, labeled **E**, **D**, **C**′, **C**, **B**, and **A** respectively. On the right side, these are at $f_q$=0.500, 0.506, 0.512, 0.513, 0.517, and 0.522, labeled **a**, **b**, **c**, **c**′, **d**, and **e**. These avoided crossings are labeled, and the energy levels in the double well potential above the band diagram are likewise labeled. **C** in the energy band diagram comes from the alignment of **a** (in the right well) and **C** (in the left well), while **c** in the energy band diagram comes from the alignment of energy levels **A** and **c**. Since all the alignments are between a higher energy level in the deeper well and the lowest level in the shallow well, the avoided level crossings are designated by the label of the energy level in the deeper well.

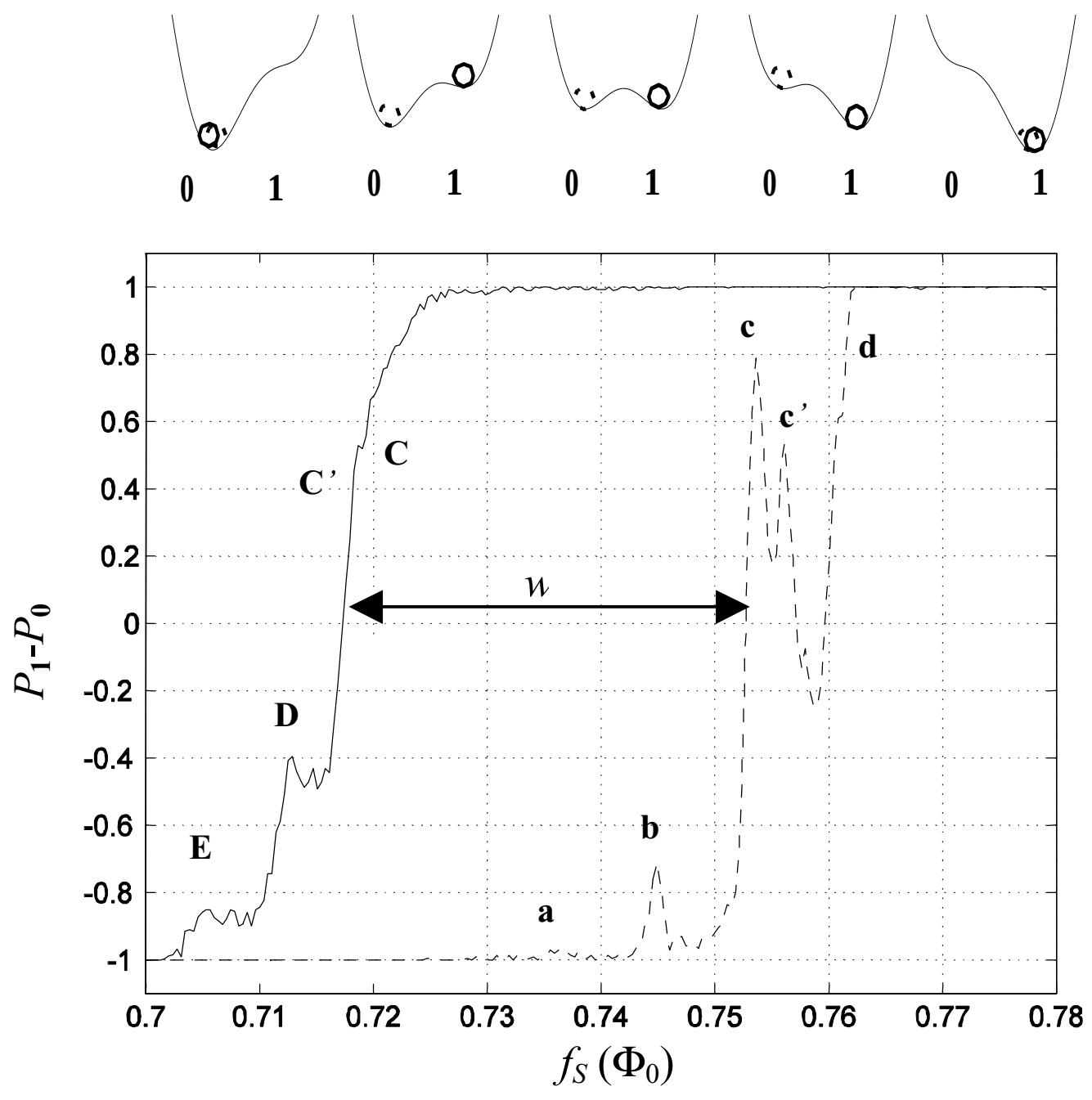

(a)

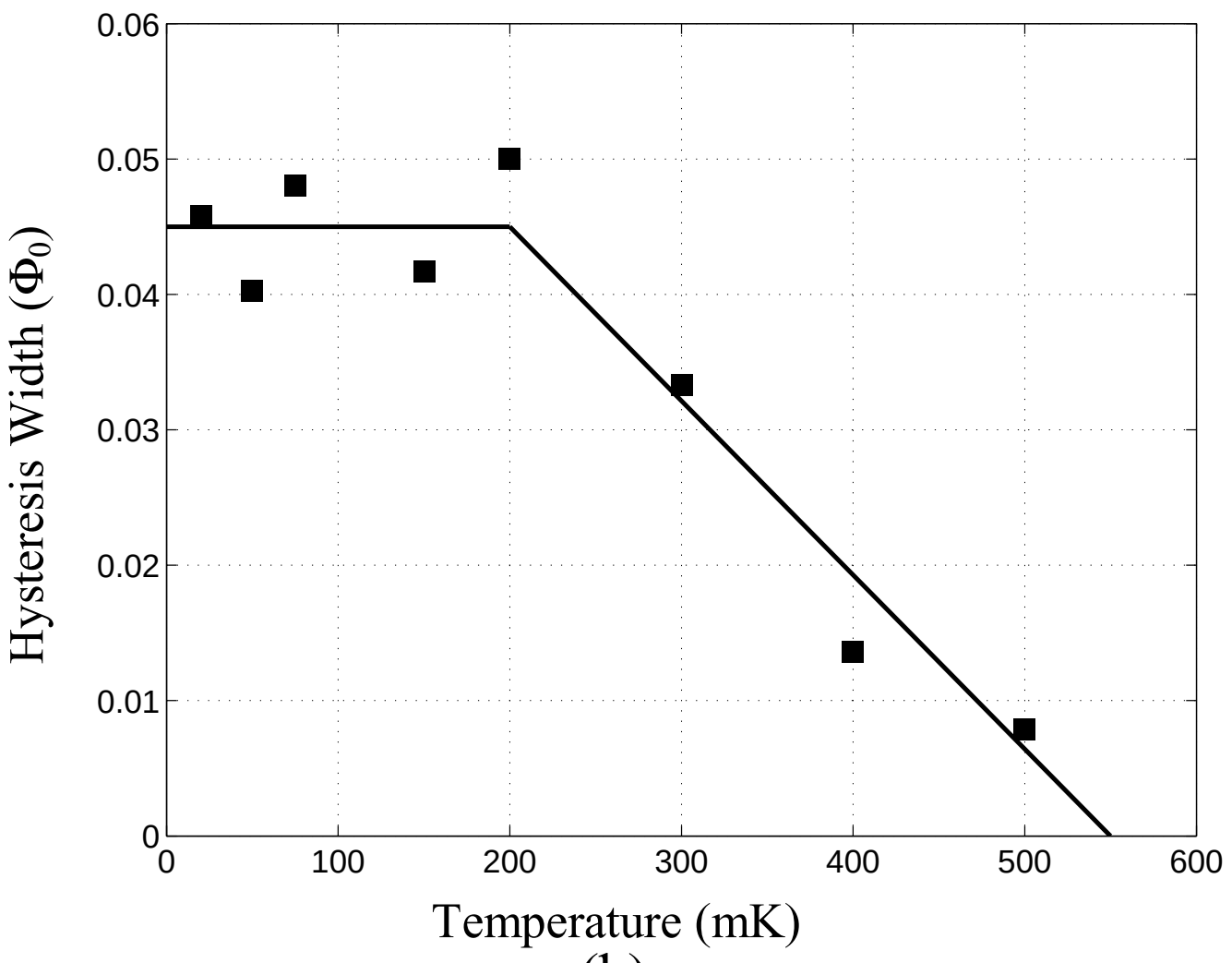

(b)

FIGURE 3. (a)The hysteresis measurement at 15 mK bath temperature. Above the figure are one-dimensional cuts of the potential that show the shape of the double-well potential at the various frustrations. The plot shows the proportion of switching events where the qubit is measured in the **1** state minus the proportion where it is found in the **0** state ($P_1$-$P_0$) against the magnetic flux bias of the dc SQUID in units of flux quanta ($f_S$). The solid line is for a qubit prepared in the **1** state, represented in the double-well diagram as a solid circle. The dashed line shows the measured qubit state when it is prepared in the **0** state, corresponding to the dashed circle in the double-well potential diagrams. The dashed line shows numerous peaks and dips, while the solid line's structure is less pronounced. Multiple scans over the same region produce the same results. The width of the hysteresis is labeled in this figure with a *w*. (b) As the temperature increases, the hysteresis loop closes. The points on this graph show the width of the hysteresis loop versus temperature. It is nearly constant for low temperatures, and nearly linear for higher temperatures. The line serves as a guide for the eyes.

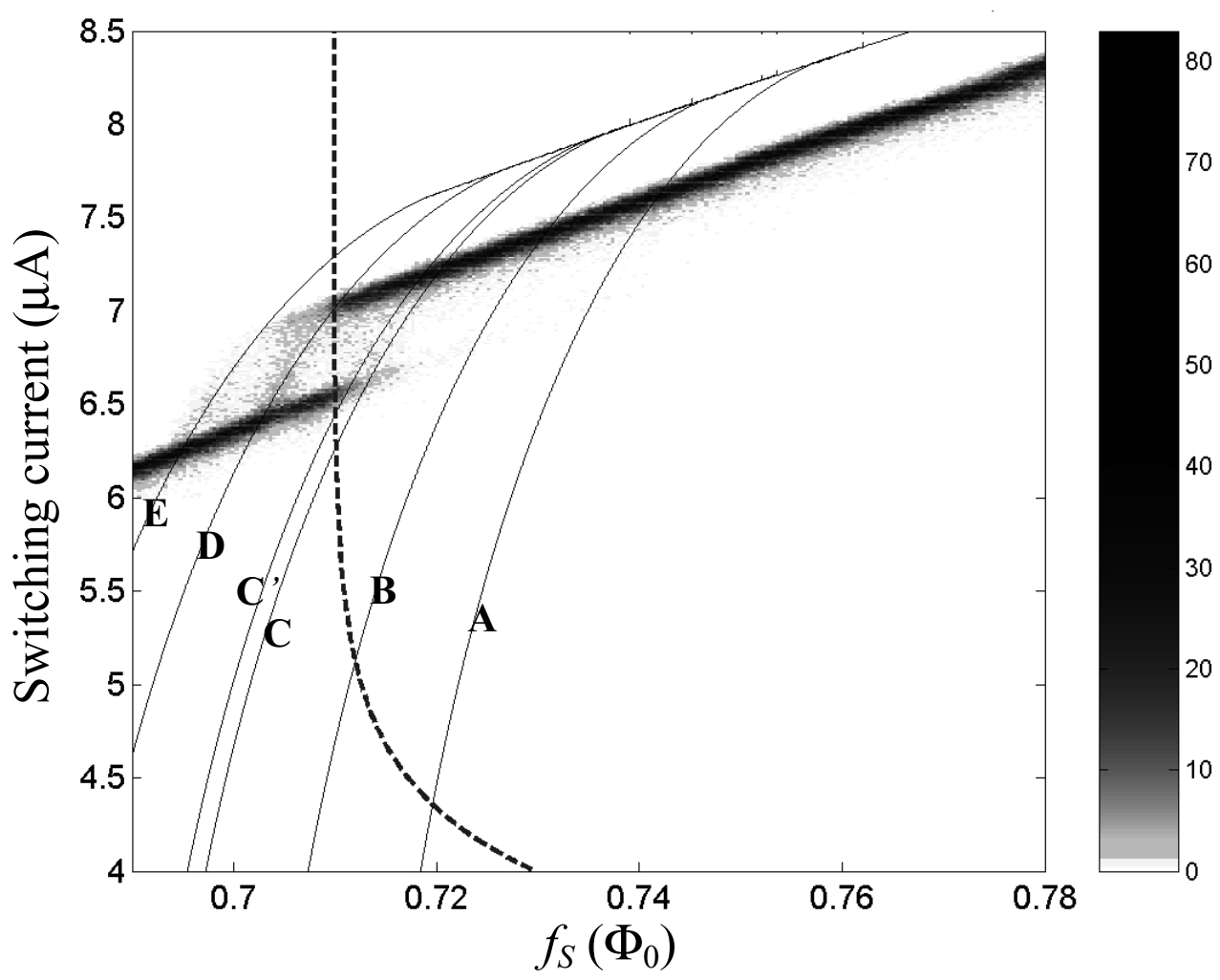

(a)

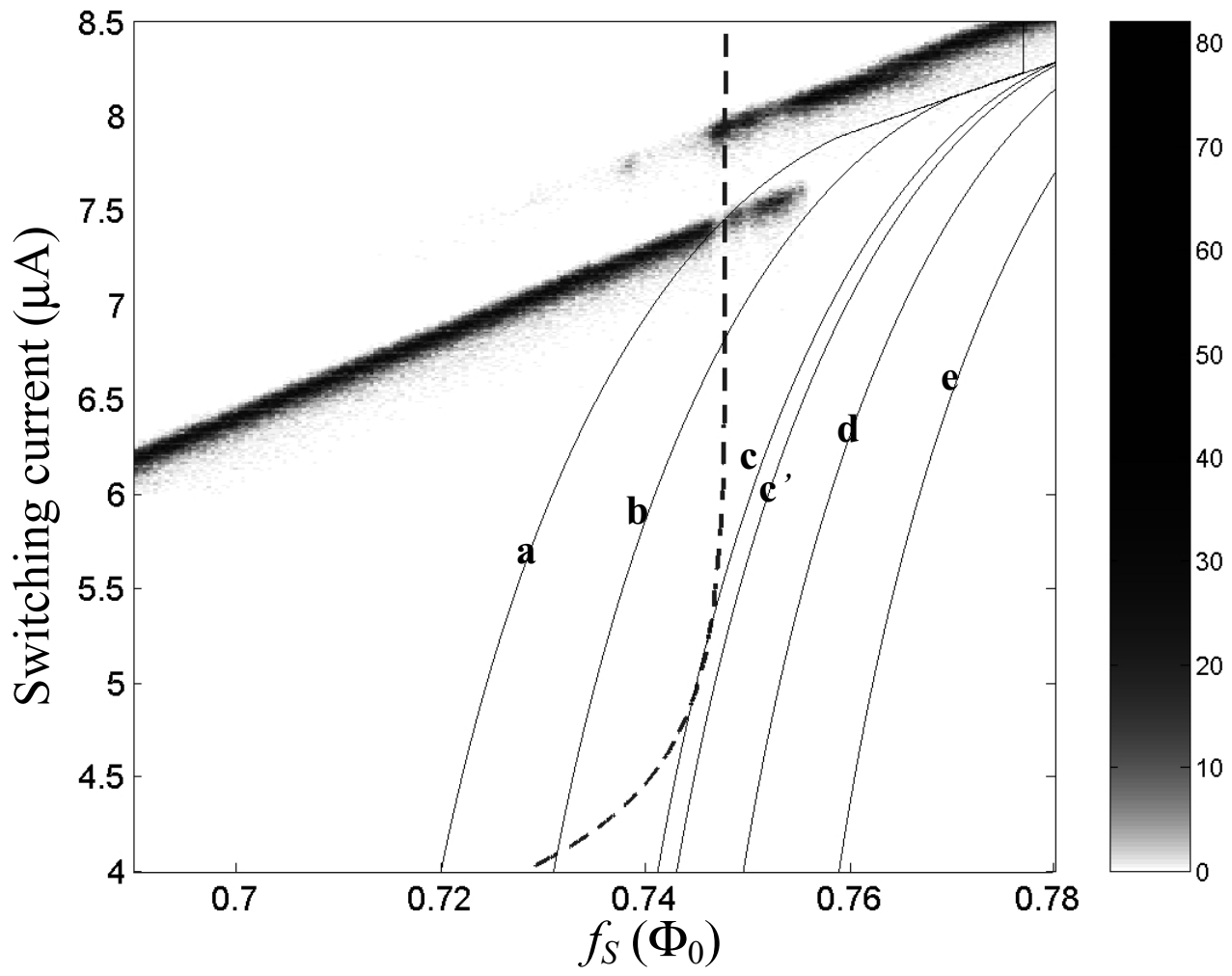

(b)

FIGURE 4. (a) A contour plot of the switching events at various external flux biases when the qubit is prepared in the **1** state. Each vertical slice is a histogram of an ensemble of switching measurements taken at a fixed external flux bias. The horizontal axis is the external flux bias of the SQUID, $f_S$. The solid lines are lines of constant $f_q$ that correspond to level crossings in the qubit labeled according to the convention in Fig. 2. The path followed when the current bias of the SQUID is ramped is not a straight vertical line, since the external flux bias is also changing due to the state preparation. The path for a representative measurement, where $f_S$=.743, is shown by the dashed line. (b) The switching events when the qubit is prepared in the **0** state. Note that the dashed line is briefly tangential with one of the solid lines.

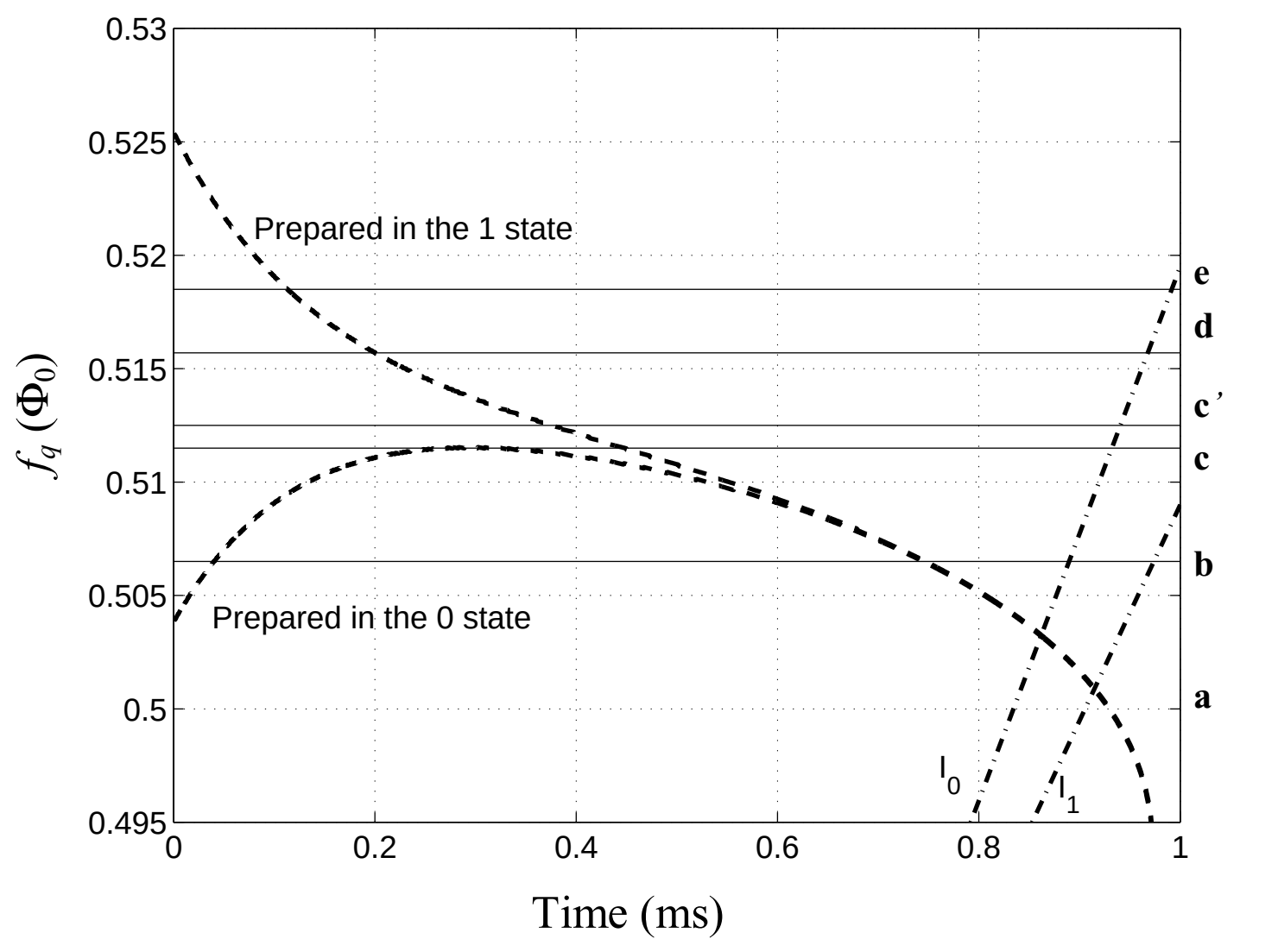

FIGURE 5. The trajectory followed by $f_q$ during the SQUID bias current ramp when $f_S$=0.743 ($f_S$=0.743 would correspond to $f_q$=0.469 if the qubit were not influenced by the circulating current in the SQUID). The solid lines are level crossings of the qubit (labeled according to the convention in Fig. 2), while the dashed lines are the paths followed when the qubit is prepared in the **0** state and in the **1** state. The plateau that occurs when the qubit is prepared in the **0** state causes sharp peaks in the data. The dash-dotted lines are the times at which the switching currents for state **0** and **1** are reached for each value of $f_q$.

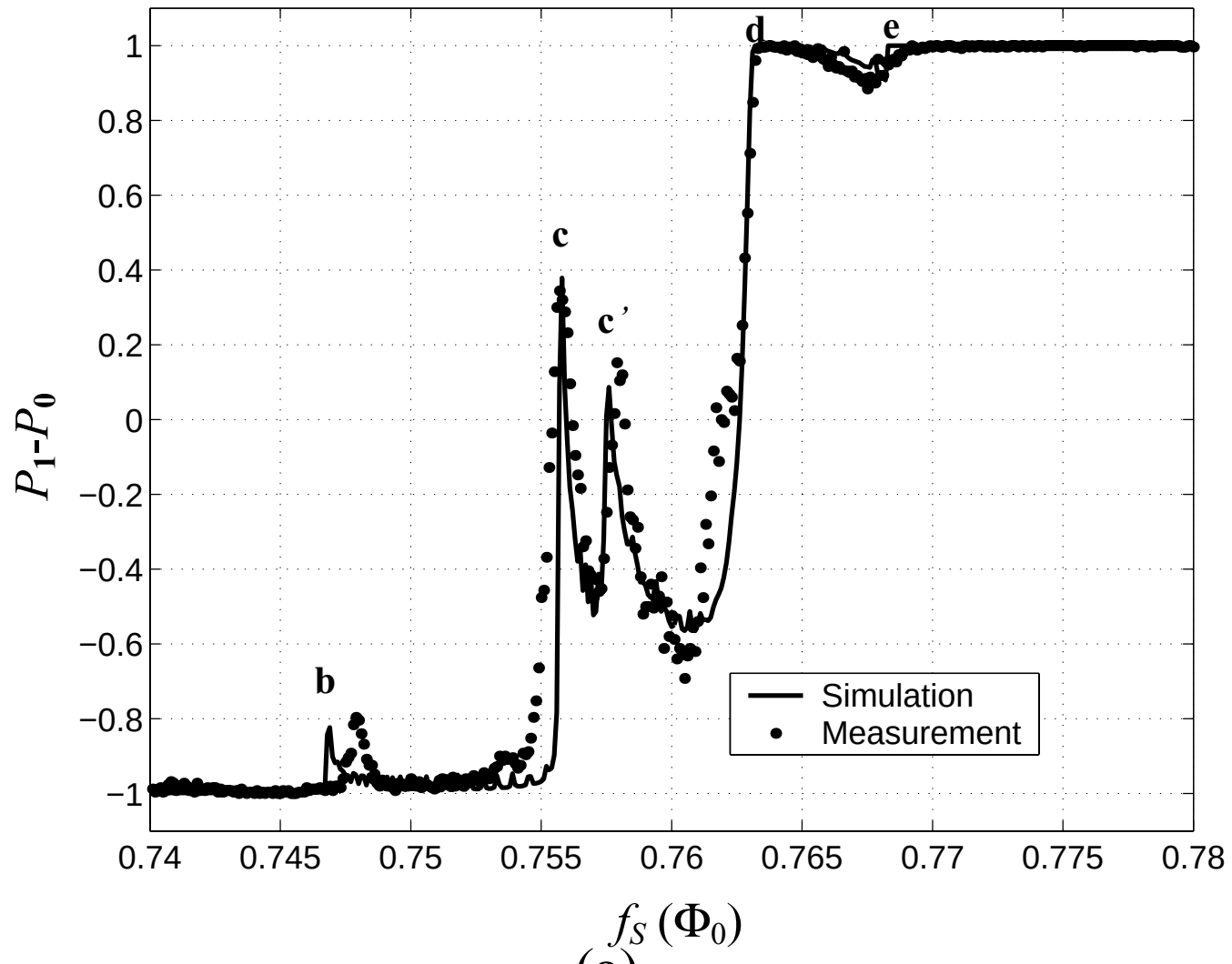

(a)

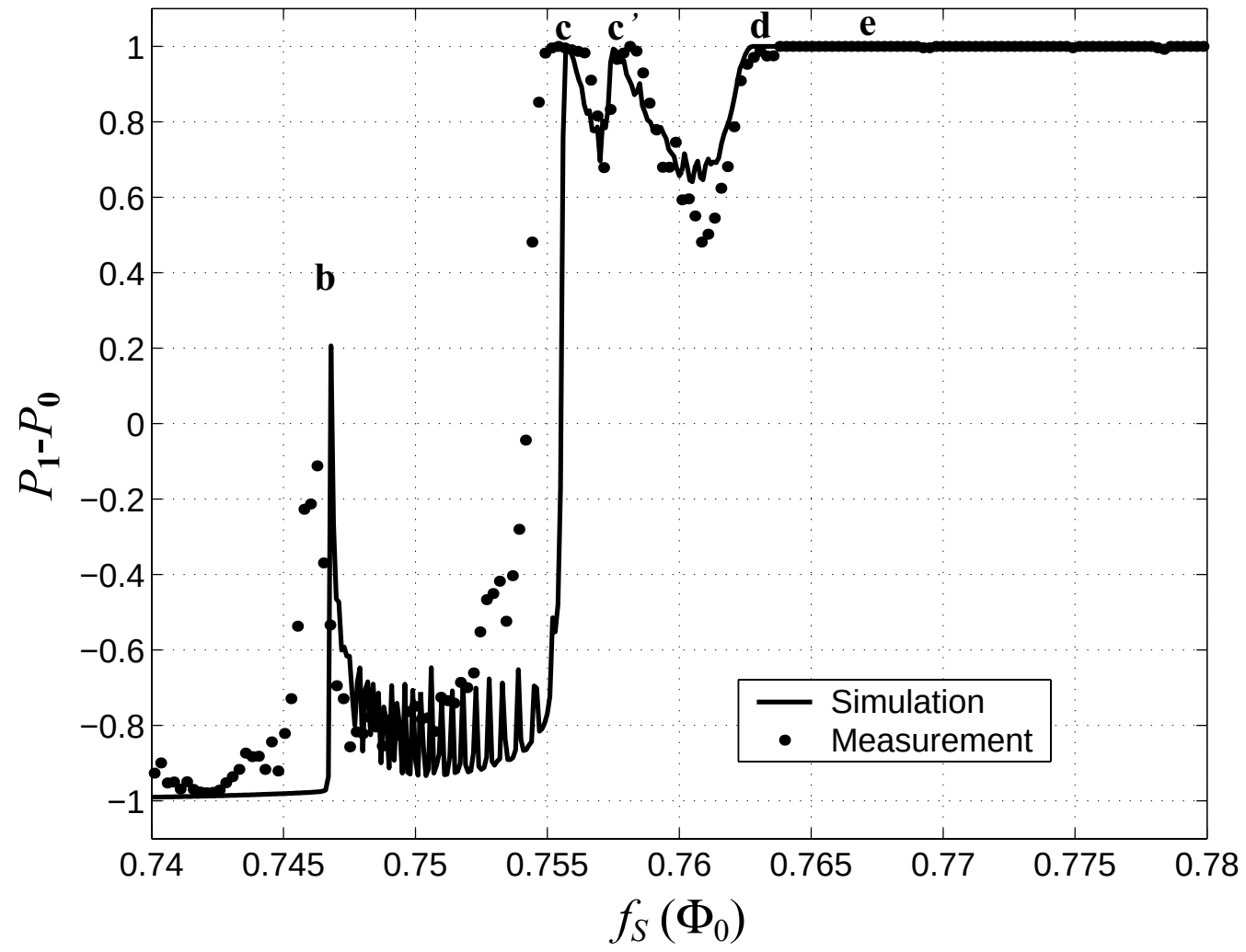

(b)

FIGURE 6. The qubit state when the SQUID is ramped at a rate of (a) 60 Hz and (b) 12.5 Hz. The solid lines correspond to the theoretical model, while the solid circles are actual data points. The two show reasonably good agreement. The slower ramp rate results in a higher probability that the qubit will transition to the **1** state, as is made clear by the growing peaks. The oscillations in (b) between $f_S$=.748 and .755 are artifacts of the numerical simulation. They decrease as resolution is increased, but resolution is limited by computer memory constraints. The peak labels correspond to the avoided crossings indicated in Fig. 2

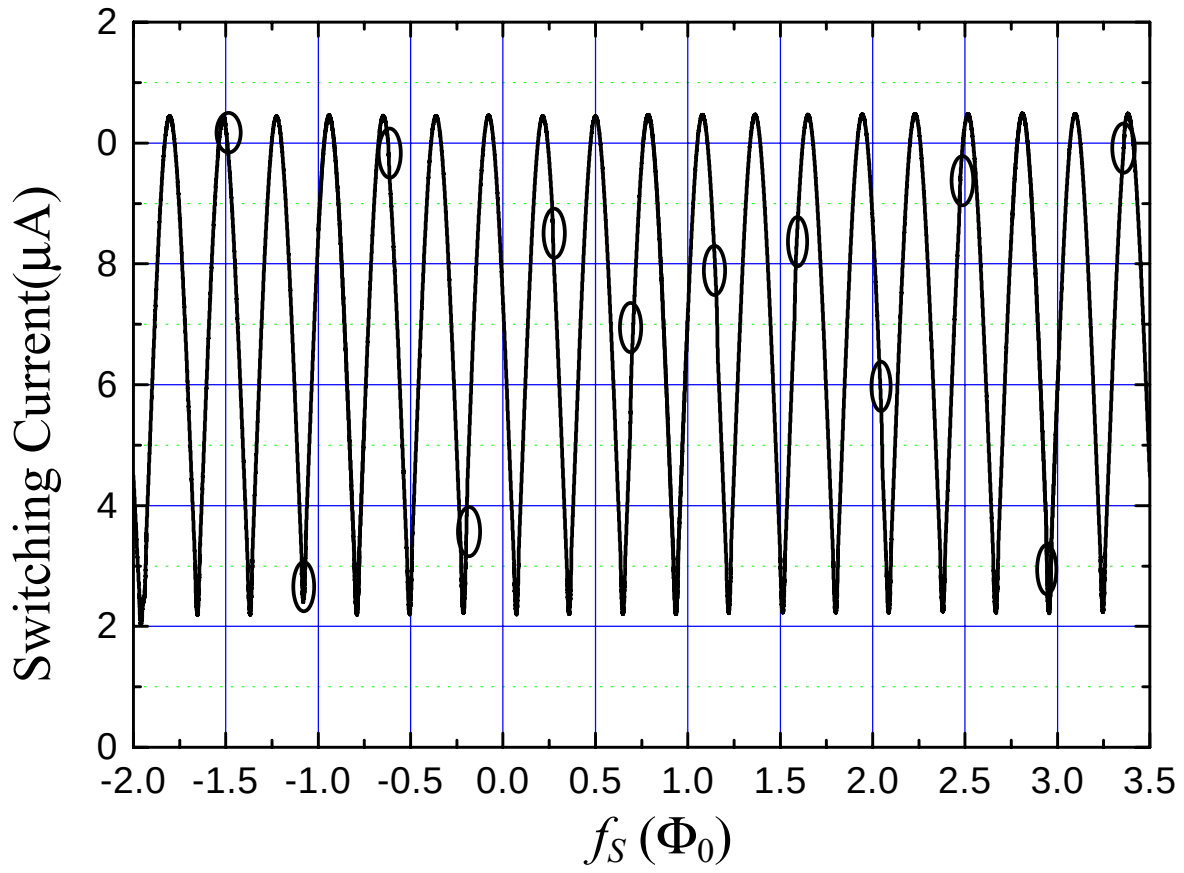

(a)

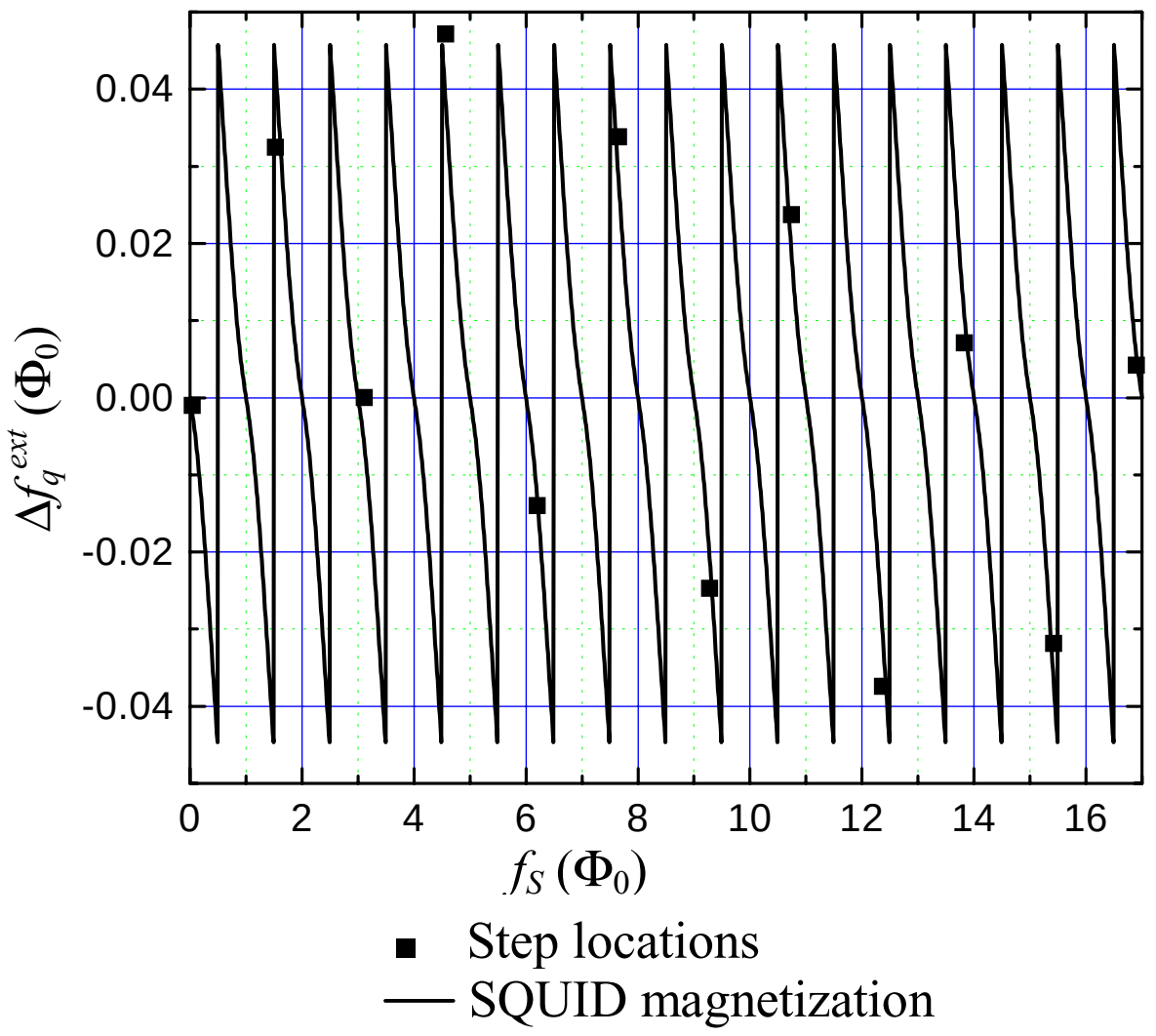

(b)

FIGURE 7. (a) This is the dc SQUID switching current curve, which approximately follows $<I_{sw}>\sim 2I_{c0}|\cos(\pi f)|$. The locations of the qubit steps are circled. The dc SQUID switching current is periodic with magnetic field, while the qubit step is nearly periodic. (b) The solid squares represent the deviations of the measured qubit step locations from a perfect periodicity of 1.53 SQUID periods per qubit period, while the solid line shows the magnetic field from the SQUID's circulating current which couples to the qubit. This is periodic with the SQUID's frustration, and accounts for the deviations from perfect periodicity.